 \documentstyle[prl,aps,epsfig]{revtex} 

\begin{document}

 \twocolumn[\hsize\textwidth\columnwidth\hsize\csname@twocolumnfalse%
 \endcsname 

\draft


\title{ \vskip -0.5cm
         \hfill\hfil{\rm\normalsize Printed on \today}\\
The Ratio Problem in Single Carbon Nanotube Fluorescence
Spectroscopy }

\author{ C. L. Kane
and E. J. Mele}

\address{Department of Physics, Laboratory for Research on the
Structure of Matter,\\
University of Pennsylvania, Philadelphia, Pennsylvania 19104}

\date{Received \hspace{3.0cm}}

\maketitle


\begin{abstract}  The electronic bandgaps measured in fluorescence
spectroscopy on individual single wall carbon nanotubes isolated
within micelles show signficant deviations from the predictions
of one electron band theory. We resolve this problem by developing
a theory of the electron hole interaction in the photoexcited
states. The one dimensional character and tubular structure
introduce a novel relaxation pathway for carriers photoexcited
above the fundamental band edge. Analytic expression for the
energies and lineshapes of higher subband excitons are derived,
and a comparison with experiment is used to extract the value of
the screened electron hole interaction.
\end{abstract}




 ] 

\narrowtext  By isolating single wall carbon nanotubes (SWNT's) in
micelles O'Connell {\it et al.} \cite{rice} have observed
midinfrared fluorescence from individual SWNT's in solution.
Measurements of the photoluminescence efficiency (PLE) as a
function of the exciting optical frequency provide a useful probe
of the optical excitations of individual SWNT's. These are of
fundamental interest and provide a basis for applications of
SWNT's as optical materials.

The spectra obtained in these experiments contain features not
anticipated by existing models for the nanotube electronic
structure \cite{dress,ham,mint,sait}. In this Letter we study the
``ratio problem"-- the ratio between critical energies measured
in these experiments violates a robust, model independent
prediction of conventional one electron theory. Here we show that
the ratio problem is resolved by identifying the effects of the
electron hole interactions in the optically excited states. The
magnitude of this problem can be used to extract an experimental
assignment of the strength of the screened Coulomb interactions
between the excited electron and hole. We develop a model for the
final state interactions following optical excitations into the
higher subbands, and find that it contains several unique
intrinsic relaxation mechanisms that derive from one
dimensionality and tubular geometry of SWNT's. We estimate the
intrinsic lifetime and derive an analytic expression for the
lineshape due to these processes.

The experiments of O'Connell {\it et al.} \cite{rice} measure
absorption and fluorescence in solutions that contain micelles
enapsulating single tubes of various radii and chiralities. The
experimental situation is sketched in the left panel of Figure 1,
where optical excitation of an electron hole pair into a high
energy azimuthal subband (a) is followed by decay into the lowest
subband from which the pair recombines with the emission of a
photon (b). The band gaps depend on the tube radius and
chirality.  By measuring the dependence of the fluorescence at a
{\it single} frequency on the exciting frequency, one isolates
from the ensemble the electronic excitations of one (or possibly a
few) tube wrappings.  Thus, peaks in the PLE can be assigned to
electronic transitions between the quantized subbands of
individual tubes.

The ratio of the frequency of the {\it second} peak in the PLE to
the fluorescence frequency measures the ratio of the critical
energies for excitations between the first two pairs of quantized
azimuthal subbands of a single tube, as shown in Figure 1.
Crudely, the ratio of the second absorption frequency to the
emission frequency should be two because of the quantization of
the azimuthal component of the crystal momentum
\cite{dress,ham,mint,sait}. This prediction is not exact and is
violated by effects arising from the tube curvature and  from the
threefold anisotropy  (the so called ``trigonal warping") in the
band structure \cite{rice2}. However these effects depend on the
tube radius and vanish for large radius tubes. Thus one electron
theory predicts that the frequency ratio must {\it
asymptotically} approach two in the limit of large tube radius.
Experimentally it does not, and instead approaches an asymptotic
value nearer to 1.75. This is significant for the optical
properties, and corresponds to a shift of the expected locations
of  band edges as large as $2000 \, {\rm cm^{-1}}$! We refer to
this problem as the ``ratio problem."

The ``ratio problem" can be identified unambiguously by studying
the aymptotic gap ratio in the limit of {\it large radius} tubes;
the observed ratios for small radius tubes are sensitive to tube
curvature and band structure anisotropy. However, even for small
radius tubes the band model fails to describe the data since it
predicts gap ratios larger than two for approximately half of the
small radius tubes measured, contrary to experiment. Likewise,
quasiparticle self energy effects in the {\it single particle}
Green's function do not resolve the ratio problem for large tube
radii since the self energy can be linearized in wavector $q$
close to the zone corner. Therefore the quasiparticle self energy
corrections also satisfy the aymptotic ``ratio equals two" rule
for large radius tubes. As we show below the ratio problem can be
traced to interaction effects in the electron-hole Green's
function that persist in the large radius limit.

Previous measurements of the electronic spectra using single
electron tunneling \cite{wild,odom} have been interpreted within
a conventional band picture including the effects of trigonal
warping. However, unlike tunneling spectroscopy, optical
excitation produces {\it two} charged carriers: the excited
electron and its valence band hole which are bound by their
Coulomb attraction \cite{ando} (Figure 2(a),(c)) producing an
excitonic bound state. Our theory uses a variational method to
study the long range interactions (Figure 2(a)) that bind the
exciton.  For a nanotube of radius $R$ embedded in a medium with
dielectric constant $\kappa$ it is convenient to introduce the
energy parameter $\Delta = \hbar v_F/R$ where $v_F$ is the Fermi
velocity, and the dimensionless coupling constant $\tilde
{\alpha} = \alpha/\kappa = e^2/(2 \pi \kappa \hbar v_F)$. All
energies in the problem can be expressed in units of $\Delta$ and
$\tilde {\alpha}$ characterizes the strength of the screened
Coulomb interaction. Our variational ansatz for the pair
wavefunction for the $n$-th subband exciton as a function of the
relative coordinate $\Phi_n(\delta z) \sim \exp(-|\delta
z|/\xi_n)$ gives an estimate of its localization length $\xi_n
\approx 1.5  \kappa R/  n$ and binding energy $\delta E_n =   n
\alpha \Delta /1.5 \kappa^2$. Note that this binding energy
scales with the size of the unrenormalized gap $2n \Delta/ 3$ and
reduces its magnitude by $\approx 10 \%$. For reasons discussed
below, this result of the variational theory disagrees with the
result of Ando's analysis of the exciton using a screened Hartree
Fock theory of the interaction \cite{ando}. While the electron
hole interaction reduces the observed threshholds for interband
transitions, the binding energy scales with the subband index $n$
and therefore this does not change the observed {\it ratios} of
the intersubband transition energies. Since the width of this
bound state can be much larger than a lattice constant we study
the relaxation of the exciton in the Wannier limit.

\begin{figure}[htbp]
\epsfxsize=3.5in
         \centerline{\epsfbox[92 438 527 698]{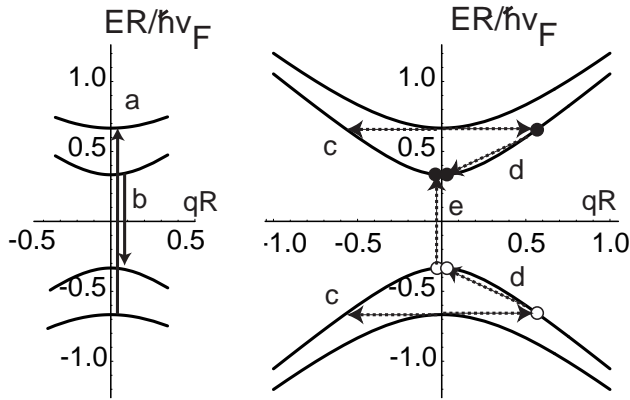}}
         \vspace{10mm}
\caption{Left: the excitation pathway leading to the ratio
problem in SWNT fluorescence spectroscopy. Electron hole pairs
excited into the second azimuthal subband (a) relax to the first
azimuthal subband where they radiatively recombine (b). Right: the
electron hole pair in the second subband can decay into a  single
electron hole pair in the first subband (c), or to a two
electron-two hole state (d-e)} \label{FIG1}
\end{figure}

Additionally, the Coulomb interaction mediates finite momentum
{\it interband} relaxation processes that conserve the total
crystal momentum (Figure 2(b)).  These have an unusual spectral
structure because of the one dimensionality of the nanotube and
the quantization of the azimuthal component of the crystal
momentum.

An exciton created at the second subband edge can relax by
decaying into an unbound electron hole pair in the first subband
with nonzero kinetic energy (as shown in process (c) of Figure
1)). This process transfers one unit of azimuthal crystal
momentum from the electron to the hole. The decay rate can be
calculated using the Golden Rule which gives the second subband
exciton a self energy
\begin{eqnarray}
\Sigma_2(E) &=&
\lim_{\delta \rightarrow 0} \frac{\tilde {\alpha}^2 \hbar v_F}{R}
\int d \zeta \,\, \frac{|V_1(\zeta)|^2 |M_{21}(\zeta)|^2}{(E+i
\delta)/\Delta - 2
\sqrt{(1/3)^2 + \zeta^2}} \nonumber\\
  &=& i A(E)  \tilde {\alpha}^2 \Delta
\end{eqnarray}
where the momentum transferred in the scattering $q = \zeta/R$,
$V_1(\zeta) = 2 I_1(\zeta) K_1(\zeta)$ is the Coulomb kernel for
interband scattering at wavevector $\zeta$ along the tube axis
expressed in terms of the Bessel functions of imaginary argument
$I_m$ and $K_m$, $M_{21}$ is the scattering matrix element
evaluated in the effective mass theory \cite{km}.   Integrating
Equation (1) we find  $A(4 \Delta/3) \approx 2.3$.

\begin{figure}[htbp]
\epsfxsize=2.5in
         \centerline{\epsfbox[115 392 465 730]{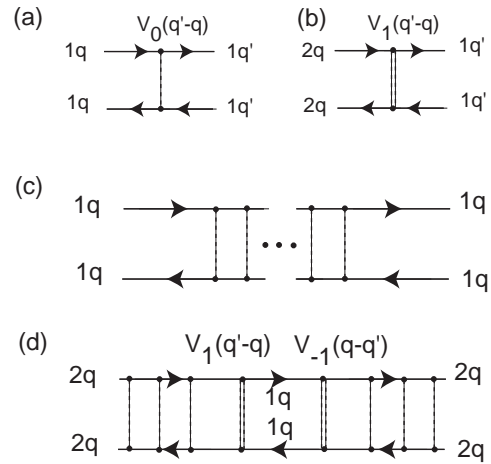}}
         \vspace{2mm}
\caption{The Coulomb interaction between the electron and hole is
described by two fundamental scattering processes: (a) intraband
scattering, conserving the azimuthal components of crystal
momenta and (b) interband scattering with finite momentum transfer
in the azimuthal direction.  Interactions denoted by the dashed
lines preserve the azimuthal crystal momentum, those denoted by
double dashed lines carry one unit of the azimuthal crystal
momentum $\hbar/R$. (c) is the multiple intraband scattering that
leads to exciton formation and (d)describes the decay of the
second subband exciton into electron hole pairs in the first
subband.} \label{FIG2}
\end{figure}

A remarkable feature of the nanotube electronic spectrum is that
the interband excitations in the second subband have sufficient
energy to decay into  {\it pairs} of first subband particle hole
excitations as depicted in Figure 1 (d-e). A typical process that
contributes to this decay is shown in Figure 3(a) where a hot
electron excites a second electron hole pair out of the vacuum.
(There are three related though inequivalent processes that are
not shown in this Figure.) The amplitudes for these processes
first appears in the perturbation theory at order $\tilde
{\alpha}^2$.

\begin{figure}[htbp]
\epsfxsize=2.5in
         \centerline{\epsfbox[78 346 536 699]{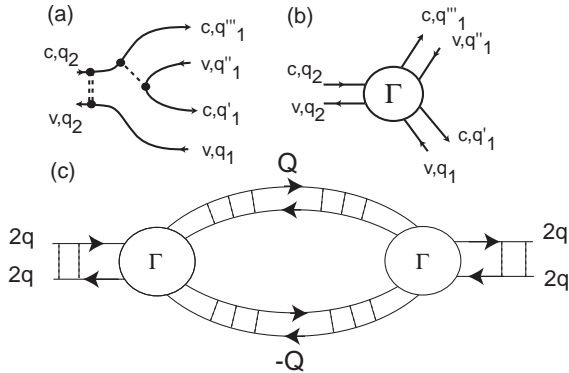}}
         \vspace{10mm}
\caption{An electron hole pair in the second azimuthal subband
can relax into a final state containing two electrons and  two
holes. (a) A typical amplitude for this relaxation process, (b)
the general vertex for the relaxation, (c) the exciton self
energy produced by this process. } \label{FIG3}
\end{figure}

All the scattering processes that contribute to this relaxation
channel are constrained by conservation of crystal momentum so
that the four momenta in the final state of Figure 3(b) sum to
zero. At the low energies relevant to our problem the situation is
further simplified by projecting this amplitude onto the final
state containing a pair of first subband excitons with momenta
$\pm Q$. The projection is carried out by integrating over the two
relative momenta of the bound electrons and holes, so that the
phase space for the final state is {\it one dimensional} indexed
by the single remaining free momentum $Q$. Thus the spectral
weight for these final states diverges at low energy proportional
to $1/\sqrt{E}$.  This leads to the unusual situation sketched in
the inset to Figure 4. The electron-hole excitation at the second
band edge resonates with a nonsingular background of
electron-hole excitations from the lowest azimuthal subband (this
is described by the self energy of Equation 1) and hybridizes with
the two electron-two hole continuum that has a {\it one
dimensional singularity} at its threshhold.

The vertex function $\Gamma(q_2,q_1,q_1',q_1'')$ for this latter
process depends only weakly on the momenta over the width
$\xi^{-1}$ relevant to describe the Wannier exciton. Using this
fact we obtain
\begin{eqnarray}
\Gamma  &=&  \frac{\tilde {\alpha}^2 \hbar v_F}{R}
\int_{-\Lambda}^{\Lambda} d \zeta \,\, \frac{V_1(\zeta) V_0(\zeta)
M_{2111}(\zeta) }{2/3 -
\sqrt{(1/3)^2 + \zeta^2}} \nonumber\\
&=&  B  \tilde {\alpha}^2 \Delta e^{i \theta}
\end{eqnarray}
where $V_m(\zeta)$ gives the Coulomb kernels for scattering with
$m$ units of crystal momentum in the azimuthal direction and with
$\delta q = \zeta/R$ along the tube axis. The factor $M$ in the
integrand is a matrix element that we evaluate using the single
particle states in the effective mass wavefunctions on the tube
\cite{km}. $\Lambda$ is an ultraviolet cutoff on the integral for
for which a Debye approximation gives the estimate $\Lambda
\approx 2.7$. The integration in Equation (3) then gives $B
\approx 5.83$ and yields for the self energy of Figure 3(c)
\begin{equation}
\Sigma_4(E) \approx  \frac{\tilde {\alpha} ^4 B^2 \Delta
}{\sqrt{(4/3)^2 -(E/\Delta)^2}}
\end{equation}

Combining the results of Equations (1) and (3) we obtain an
expression for the exciton Green's function in terms of the scaled
energy $\varepsilon =E/\Delta$
\begin{eqnarray}
{\cal G}(\varepsilon) &=&
  \frac{1}{\varepsilon  -\varepsilon_o +  i A
\tilde {\alpha}^ 2 + B^2 \tilde {\alpha}^4/
\sqrt{\varepsilon_o^2-\varepsilon^2}}
\end{eqnarray}
where $\varepsilon_o$ gives the (scaled) energy of the
unperturbed exciton. The absorption lineshape ${\cal
L}(\varepsilon) = - {\rm Im}\, {\cal G}(\varepsilon)/\pi$ is
therefore completely determined by the screened coupling constant
$\tilde {\alpha}$; in Figure 4 we plot ${\cal L}(\varepsilon)$
for three representative values of the coupling. The zero in this
lineshape denotes the locations of the threshhold energy for the
two particle-two hole continuum. Note that while the electronic
bandgaps depend on tube radius $R$, scaling approximately as
$1/R$, the Coulomb interaction has the same scaling with radius.
In particular, for large radius tubes the excitonic interactions
become weak {\it in proportion} to the scale of the band gap,
expressed as a fraction of the bandgap it approaches a nonzero
limit for large $R$.

The second term in the self energy of Equation (4) is higher
order in the small parameter $\tilde {\alpha}$, but it makes a
signficant contribution to the self energy because of its
singular energy dependence. Solving for the roots of the
denominator we find that the pole for the bare second subband
exciton is shifted to lower energy $\Delta \varepsilon = -
3^{1/3}B^{4/3}\tilde{\alpha}^{8/3}/2$ and develops a width
$\delta \varepsilon = A \tilde{\alpha}^2$. In contrast, the first
subband exciton  is unaffected by any similar relaxation process,
since it does not overlap a pair continuum, and therefore it is
unrenormalized.

In the lower panel of Figure 4 we plot the ratio of the
renormalized energy of the second subband exciton to the energy
of the first subband exciton. The dashed line gives the result of
the noninteracting theory that ignores these relaxation effects.
Experiment indicates that in the limit of large tube radii the
ratio lies in the shaded region, which is well described by a
dimensionless coupling constant $\tilde \alpha \approx 0.20$. The
{\it unscreened} coupling constant is $e^2/2\pi \hbar v_F \approx
0.42$ and so the fit indicates an effective dielectric screening
of the exciton with $\kappa \approx 2.1$.  For this coupling
constant the linewidth is $\approx 0.09 \hbar v_F/R$, giving
$\delta \lambda/\lambda \approx 0.07$ which is quite close to the
experimentally observed width in PLE \cite{rice2}. This coupling
constant places the nanotube micelle in the crossover regime
where the shift and width of the line are comparable, as shown as
the bold curve in Figure 4.

A single electron injected into the second azimuthal subband edge
has insufficient energy to relax by exciting an electron hole
pair. Thus the processes illustrated in Figure 3 for the exciton
do not occur in the single particle Green's function and should
have no effect on electronic spectra measured by tunneling
spectroscopy.

The effect of the Coulomb interaction on the optical excitations
has been studied theoretically previously by Ando \cite{ando}.
His work augments the effective mass model by introducing a
Coulomb interaction between electrons that is studied within the
Hartree Fock approximation. Ando's approach predicts an
enhancement of the single particle bandgap due to the
interactions, and the appearance of a bound exciton spectrum
within this larger gap. Note however that the effective mass
Hamiltonian is parameterized to match the low lying quasiparticle
excitations of a the reference graphene sheet and thus already
includes some self energy effects due to the Coulomb interaction.
Our model treats the excitonic effects of the  Coulomb
interaction between the electron and hole, and therefore builds
in the correct separation energy within the effective mass theory.
Processes leading to the self energy of Figure 3 occur beyond the
level of a Hartree Fock treatment of the interaction.

It is interesting that if the coupling constant $\tilde {\alpha}$
were unscreened, the renormalized second subband exciton
hybridizes so strongly with the two electron two hole continuum
that its energy becomes comparable to that of the fundamental
first subband exciton. This situation is reminscent of the known
anomalous ordering of the electronic excitations of the polyenes
\cite{ts}. There the lowest singlet excitation corresponds not to
the promotion of a single electron from the highest unoccupied
molecular orbital to the lowest unoccupied molecular orbital, but
instead to a two particle-two hole excited state.

The interaction of carbon nanotubes with visible light results in
the formation of ``hot" carriers well above the lowest band edge.
Our results demonstrate that a quantitative analysis of these
excitations requires an understanding of the dominant relaxation
mechanisms for these carriers.  We find that the excitations of
the many body system at these energies mix the interband
excitations of the independent particle model with a continuum of
primitive electron-hole excitations, and the latter are
particularly strong because of the one dimensional character of
these structures. These can now be included in quantitative
analyses of spectroscopic data on these systems.

Acknowledgements: This work was supported by the Department of
Energy under Grant DE-FG02-ER-0145118, and by the National
Science Foundation under MRSEC grant DMR-00-79909. We thank Rick
Smalley and Bruce Weisman for communicating their experimental
results prior to publication.

\begin{figure}[htbp]
\epsfxsize=2.0in
         \centerline{\epsfbox[119 144 504 752]{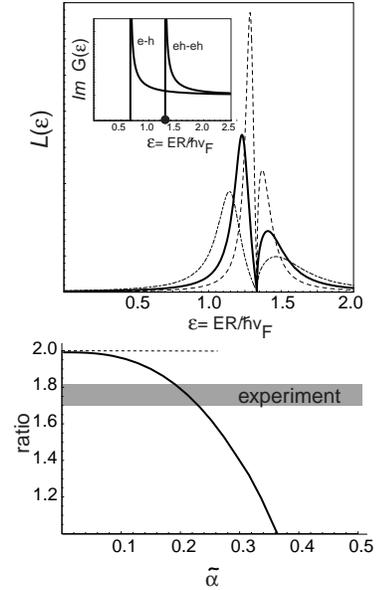}}
         \vspace{2mm}
\caption{Absorption linehapes for the second subband exciton
resonating with first subband excitations plotted as a function
of the dimensionless energy $\varepsilon=ER/\hbar v_F$.  The
inset shows the spectral densities for the electron-hole and two
electron-two hole continua. The bold dot denotes the location of
the unrenormalized second subband exciton. The curves show the
dimensionless lineshapes computed from equation (4) for three
values of the screened dimensionless coupling constant of the
theory ($\tilde {\alpha}$ = 0.15 (dashed), 0.20 (solid), 0.25
(dot-dashed)). The lower panel gives the ratio of the energy of
the renomalized exciton to the energy of the fluorescence. The
unrenomalized theory gives the dashed line and the shaded region
is obtained from the experiments of reference 2.} \label{FIG4}
\end{figure}

\end{document}